\documentstyle[preprint,aps,graphicx]{revtex}
\tightenlines
\begin{document}
\title{Thermodynamic Study of Excitations in a 3D Spin Liquid}
\author{Y.K.\ Tsui, J.\ Snyder, and P.\ Schiffer}
\address{
Department of Physics, Pennsylvania State University, University Park, PA 16802.}
\maketitle

\begin{abstract}

In order to characterize thermal excitations in a frustrated spin liquid, we have examined the magnetothermodynamics of a model geometrically frustrated magnet.  Our data demonstrate a crossover in the nature of the spin excitations between the spin liquid phase and the high-temperature paramagnetic state. The temperature dependence of both the specific heat and magnetization in the spin liquid phase can be fit within a simple model which assumes that the spin excitations have a gapped quadratic dispersion relation.

\vspace{0.1in}
\noindent PACS numbers: 75.30.Kz, 75.50.Ee
\end{abstract}

\newpage
In recent years, there has been increasing interest in the properties of spin liquid states in which the spins are highly correlated but have strong fluctuations in the low temperature limit.  These states are an important component of the physics in a variety of systems including low dimensional magnets \cite{1dliq}, cuprate superconductors \cite{cupliq}, and geometrically frustrated antiferromagnetic materials in which the topology of the lattice leads to frustration of the antiferromagnetic exchange interactions \cite{Rami94}. In these geometrically frustrated magnets, the spins do not order except at
temperatures well below the Curie-Weiss temperature
($\Theta_{\mathrm{cw}}$), and in the low temperature limit they can exhibit a variety of novel correlated spin states including the spin liquids in which the spins fluctuate within low-moment locally correlated spin clusters  \cite{clusters}.  These frustrated spin liquids have been the subject of extensive theoretical work \cite{Spinliqtheor}, and their existence has been confirmed experimentally in several materials of different frustrated topologies \cite{Rami90,Schi94,harris,Petr97,Ball96,Tsui99,Gard99,RaHeWi00,Duns00,shiga}.  The basic thermodynamic properties of frustrated spin liquids are, however, only beginning to be explored \cite{Spinliqtheor,Ball96,Gard99,RaHeWi00}, and there is neither a clear understanding of the nature of the thermal excitations at low temperatures nor a clear distinction between the expected excitations in spin liquids and those in paramagnetic states at higher temperatures.

In this paper we examine the excitations in a three-dimensional frustrated spin liquid through measurements of the specific heat $(C)$ and static
magnetization $(M)$ in a model geometrically frustrated magnet.  The field dependence of $C(T)$ suggests a crossover in the nature
of spin excitations between the spin liquid phase and the high-temperature
paramagnetic phase. Moreover, both $C(T)$ and $M(T)$ at low temperatures can be 
quantitatively described by a spin-wave-like model of excitations with a gapped quadratic dispersion relation.  The gap grows linearly with magnetic field, but appears to have a finite value even in the extrapolation to zero field.

\begin{figure}
\includegraphics[height=6in,width=5in,angle=270]{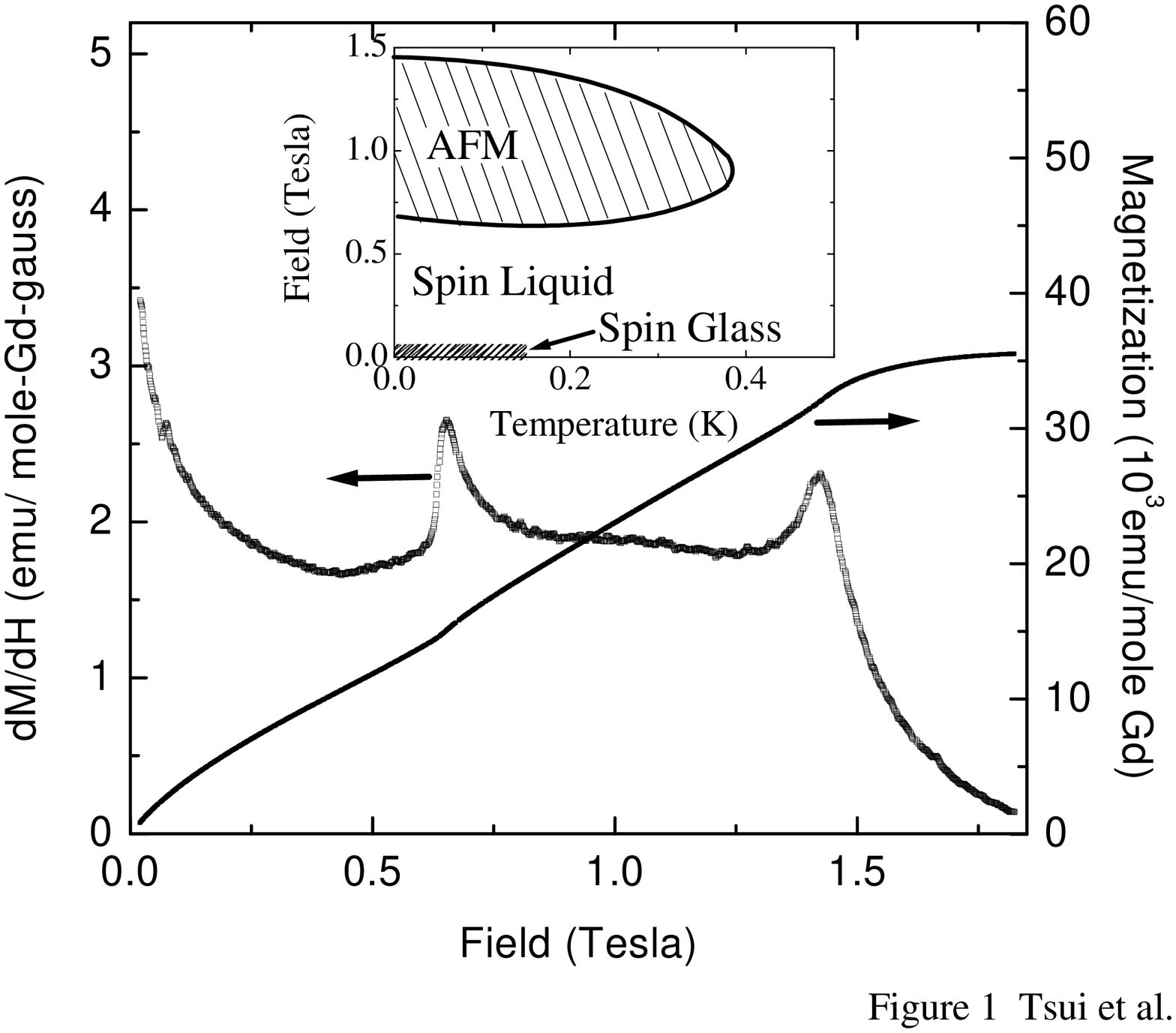}
\vspace{1.5in}
\caption{The field dependence of static magnetization and the magnetic susceptibility $dM/dH$ at
0.1 K.
The inset shows the low-temperature phase diagram of GGG including
the spin glass, spin liquid, and antiferromagnetic long-range
ordered phases [3,9].}
\label{fig:Chidc}
\end{figure}

We studied the model geometrically frustrated magnet Gd$_3$Ga$_5$O$_{12}$ (gadolinium gallium garnet or GGG) 
\cite{Schi94,Petr97,Tsui99,Duns00,KiWo79,Hov80} in which the magnetic Gd ions are located on two 
interpenetrating corner-sharing triangular sublattices within the garnet 
structure.  This geometry frustrates the antiferromagnetic 
nearest neighbor exchange interactions ($|\Theta_{\mathrm{cw}}| \sim$ 2K)\cite{KiWo79,Wolf62} and yields a 
highly unusual low-temperature phase diagram as shown in the inset to figure 
\ref{fig:Chidc} \cite{Schi94,Tsui99,Hov80}. At low fields ($H \lesssim 0.08$ T) previous studies show that the ground state of GGG is a spin glass with $T_g \sim $ 0.13 K \cite{Schi94}. At fields between 0.7 and 1.4 T the
ground state is a field-induced long-range-ordered antiferromagnetic 
state with $T_N$(max) $\sim$ 0.35 K \cite{Schi94,Petr97,Tsui99,Hov80}.
The ground state at intermediate fields, $0.08 \lesssim H \lesssim 0.7$ T, shows 
no evidence of any sort of spin ordering upon cooling from the high-temperature
paramagnetic state down to the lowest temperatures which have been studied
($T < |\Theta_{\mathrm{cw}}|/40$), and is apparently a homogeneous 
three-dimensional spin liquid phase in which the spins are strongly fluctuating even at the lowest temperatures \cite{Schi94,Tsui99,Duns00}.   

We measured $M$ and $C$ on samples of GGG cut from the same single crystal grown
by the Czochralski method.
Magnetization measurements were performed by mounting a GGG
sample on a flexible silicon paddle and applying a magnetic field gradient 
($\lesssim$ 1 T/m) in addition to a spatially homogeneous magnetic field parallel to the [100] direction.
The resulting force on the magnetized sample was proportional to the
magnetization, $F = M\partial H/\partial z$, and was measured from the 
deflection of the paddle by a capacitive technique \cite{Schi94,Saka94}. The 
sample, the magnetometer, and the sample thermometers
were immersed with liquid helium and thermally linked to the mixing chamber of a 
dilution refrigerator. As a result, we could ensure the sample was in thermal 
equilibrium with the refrigerator during the measurements. Details of the 
experimental setup and techniques for heat capacity measurements can be found in 
Ref. \cite{Tsui99}.

\begin{figure}
\vspace{1in}
\includegraphics[height=6in,width=5in,angle=270]{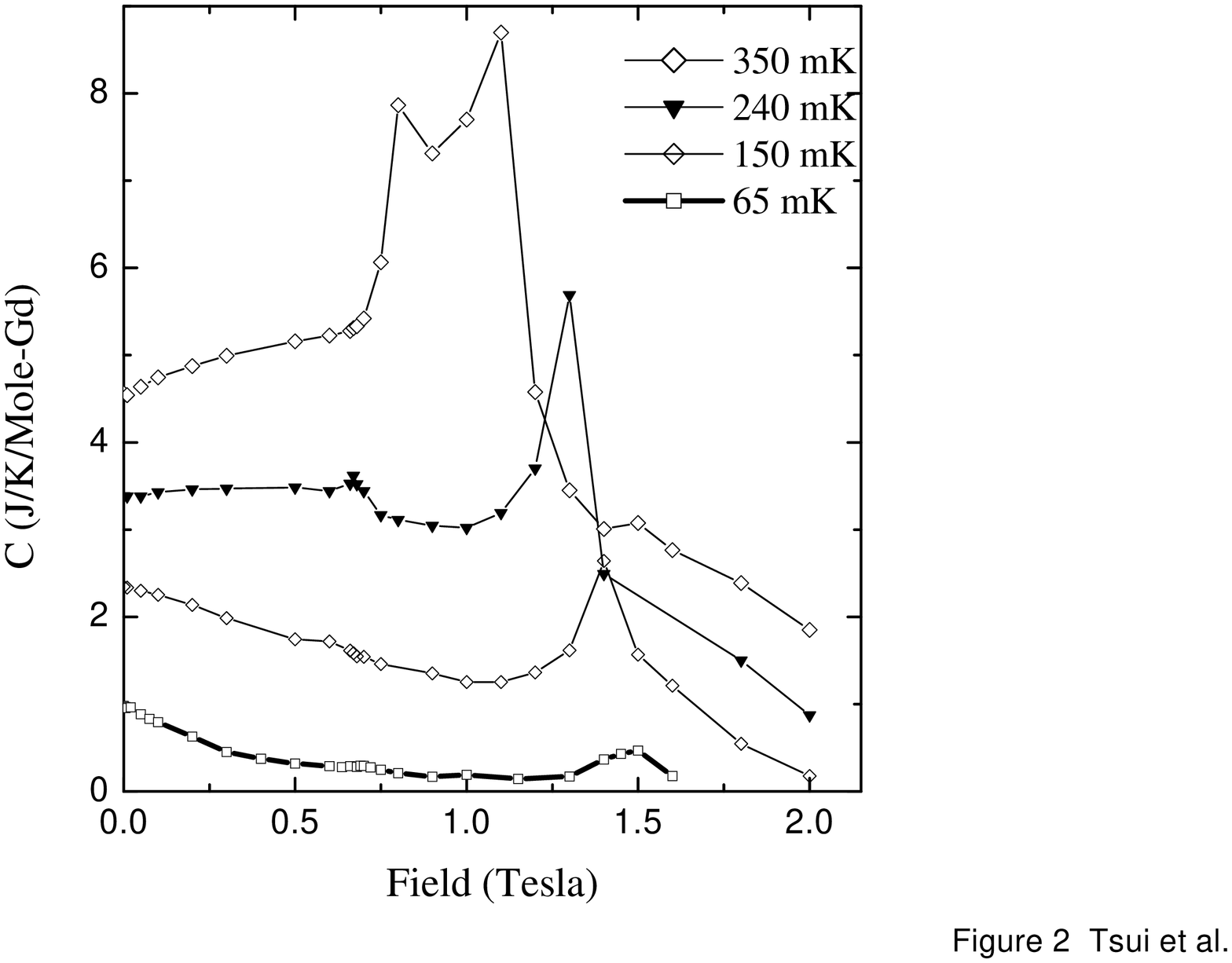}
\vspace{1in}
\caption{ Specific heat ($C$) of GGG versus applied field at several
temperatures,
note $\partial C_p/\partial T$ changes sign at $\sim$ 0.235 K).}
\label{fig:CH}
\end{figure}

In figures \ref{fig:Chidc} and \ref{fig:CH} we plot $M(H)$ and $C(H)$, and in 
both data sets there are features for the boundaries of the 
antiferromagnetic phase (clearly visible as sharp peaks in $\partial M/\partial 
H$ in agreement with previous a.c. susceptibility data \cite{Schi94} and recent Monte Carlo results \cite{Petr97}). Both $M$  and $C$ change monotonically with field in the spin liquid phase ($0.08 \lesssim H \lesssim 0.7$ T), with M(H) increasing almost linearly with field at low temperatures.  We note that $C(H)$ is only weakly field-dependent in the spin liquid phase at low temperatures, 
changing in magnitude by only $\sim$ 30 \% when the applied field is doubled 
from 0.3 to 0.6 T at $T =$ 0.065 K. Taking the spin 7/2 of the 
Gd$^{\mathrm{3+}}$ ions to set the energy scale, the $\Delta H$ of 0.3 T is 
quite large, equivalent to $\Delta T \sim 0.3\mu_{\mathrm{Gd}}/k_{\mathrm{B}} 
\sim 1.4$ K, which is
comparable to the exchange energy between spins ($|\Theta_{\mathrm{cw}}| \sim 2$ 
K). This weak field dependence of $C(H)$ is similar to that of another geometrically frustrated magnet with a spin liquid state,
SrCr$_{\mathrm{9p}}$Ga$_{\mathrm{12-9p}}$O$_{19}$ \cite{energyscale}, and is theoretically expected for a spin liquid in which the excitations are among correlated spin-cluster singlets \cite{Spinliqtheor,RaHeWi00}.

\begin{figure}
\vspace{1in}
\includegraphics[height=6in,width=5in,angle=270]{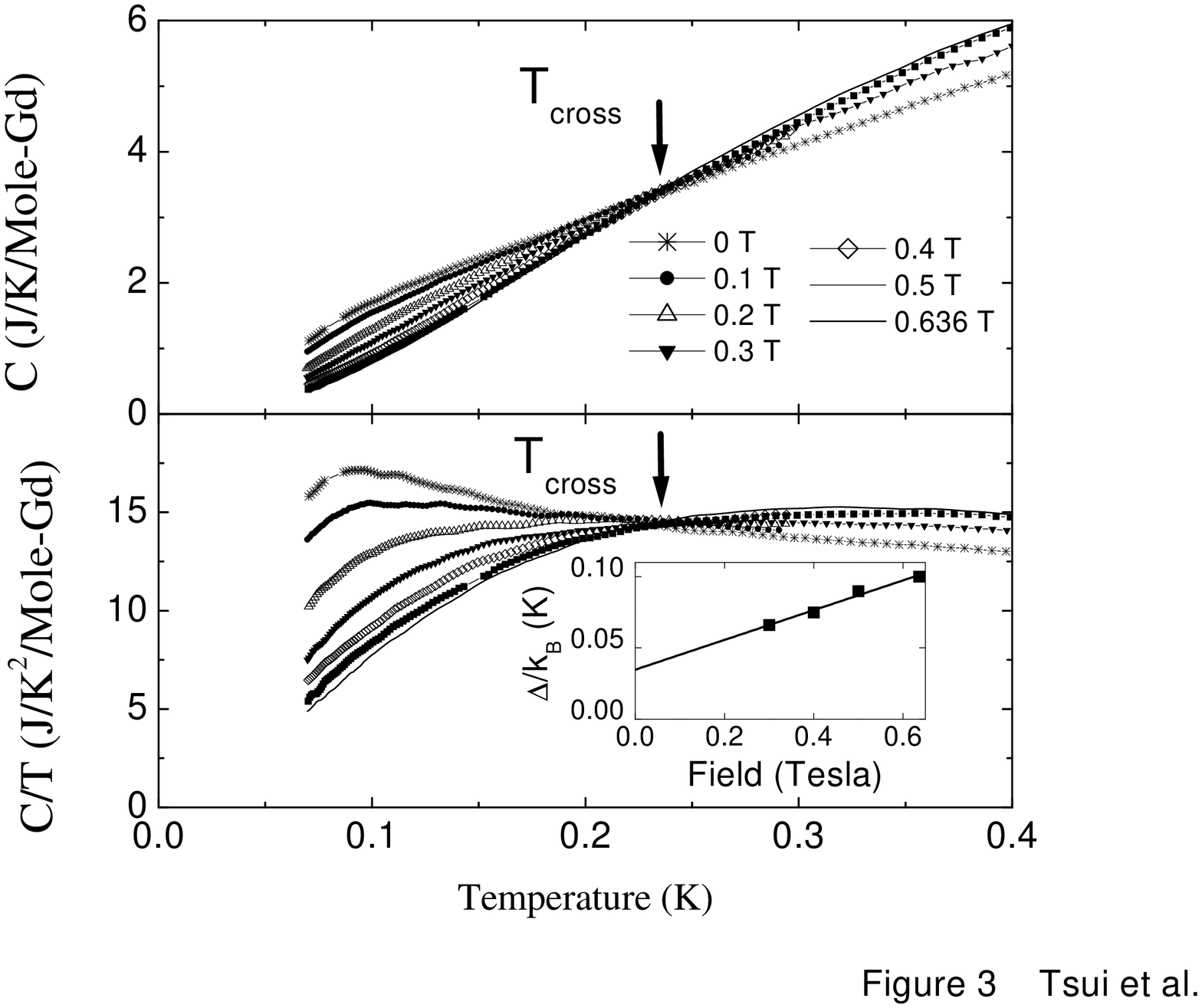}
\vspace{1in}
\caption{The temperature dependence of $C_p$ and $C_p/T$ in the spin liquid regime in different applied fields.  Note that all the curves cross at T =  $T_{cross} \sim$ 0.235 K.  The inset shows the energy gap ($\Delta$) obtained from fits to the low temperature $C(T)$ data}
\label{fig:CT}
\end{figure}

\begin{figure}
\vspace{1in}
\includegraphics[height=6in,width=5in,angle=270]{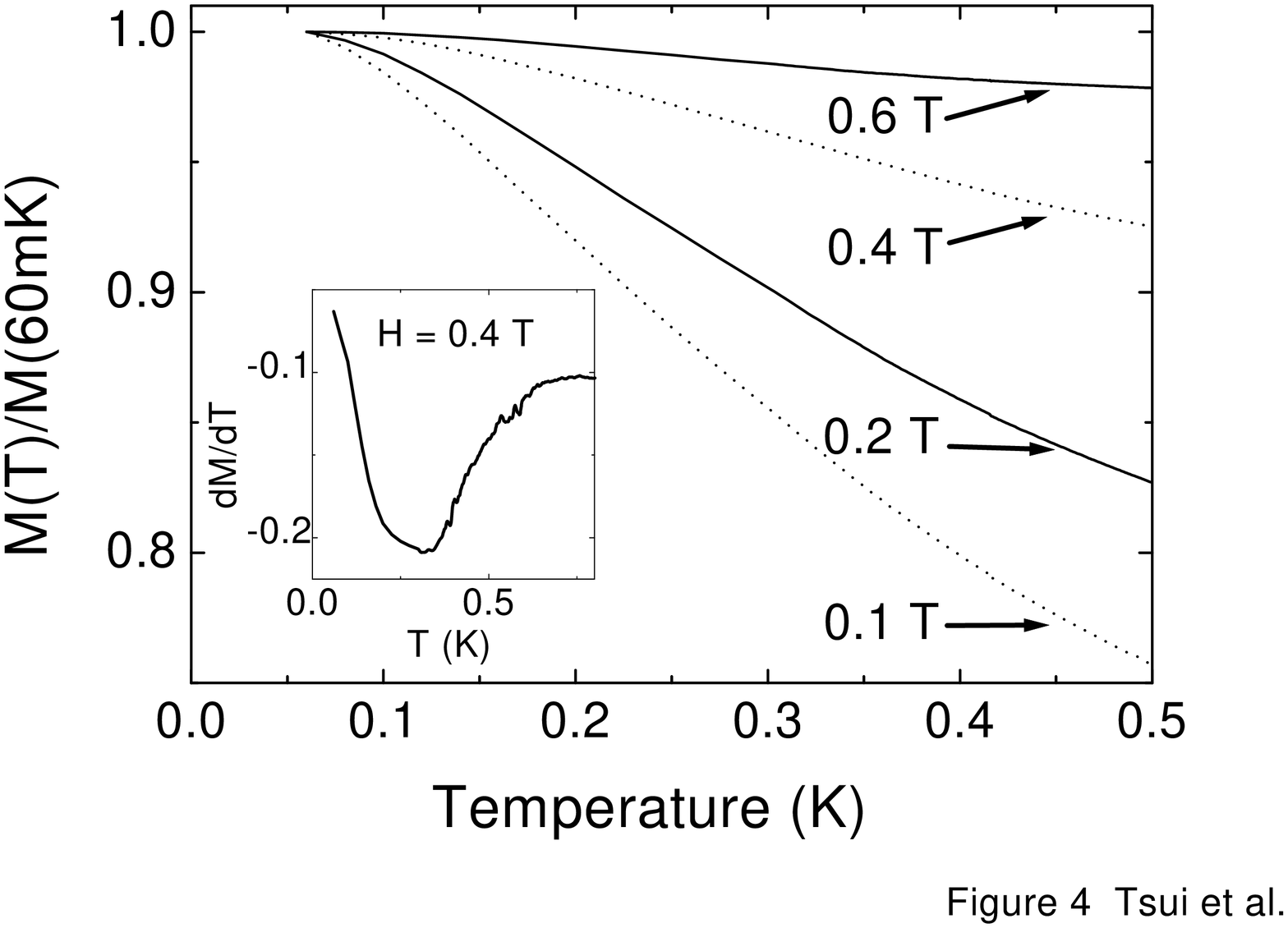}
\vspace{1in}
\caption{The temperature dependence of $M$ (normalized to 0.06 K)
in the spin liquid regime. The inset shows $dM/dT$ versus temperature, notice the
minimum at $\sim$ 0.3 K corresponding to $T_{cross}$}
\label{fig:MT}
\end{figure}

An unresolved issue in the study of geometrically frustrated spin liquids is 
whether the spin liquid phases can be differentiated from the high-temperature paramagnetic 
states since there are no sharp features in the temperature dependence of physical  properties to indicate a transition to a different phase.  This is the case for the spin liquid phase of GGG, but the specific heat data do show a distinct crossover in behavior at $T_{cross} \sim$ 0.235 K, where for all fields less than 0.7 T, the $C(T)$ curves cross (figure  \ref{fig:CT}).  This 
crossing indicates that all field derivatives of the specific heat vanish at 
$T_{cross}$, which in turn implies that $\partial C(H)/\partial H$ changes sign at $T_{cross}$. The field independence of the heat capacity at $T_{cross}$ is also evident in figure \ref{fig:CH} in that the 240 mK isotherm is constant for magnetic fields within the spin liquid 
phase.  The intrinsic nature of this crossover is confirmed 
by an inflection point in our $M(T)$ data (figure \ref{fig:MT}) which is 
thermodynamically equivalent to a zero in $\partial C(H)/\partial H$ (see equation \ref{eqn:SandM} below). 

This crossover temperature suggests that we might draw a distinction between the 
nature of spin excitations in the spin liquid state and those in the 
paramagnetic state at higher temperatures.  The change in the sign of $\partial 
C(H)/\partial H$ implies that the excitations are more ferromagnetic-like in 
the low temperature limit in that they are suppressed by a field,
and that the excitations are more antiferromagnetic-like at higher temperatures
in that they are enhanced by a field. In other words, at high temperatures the 
thermal fluctuations primarily dissociate the antiferromagnetic correlations 
between the spins, while at low temperatures, the antiferromagnetic correlations 
are unaffected by thermal fluctuations so the fluctuations primarily 
reduce the alignment of the spins with the magnetic field. This crossover 
may actually be a generic  feature of spin liquid states, since the same feature was 
recently observed by Ramirez et al. in SrCr$_{\mathrm{9p}}$Ga$_{\mathrm{12-
9p}}$O$_{19}$ \cite{RaHeWi00}.  These two cases suggest that a generic
distinction can be drawn between the low temperature spin liquid and the high temperature 
paramagnetic states in these systems, at least in the 
nature of the spin excitations in the presence of a magnetic field.

\begin{figure}
\vspace{1in}
\includegraphics[height=6in,width=5in,angle=270]{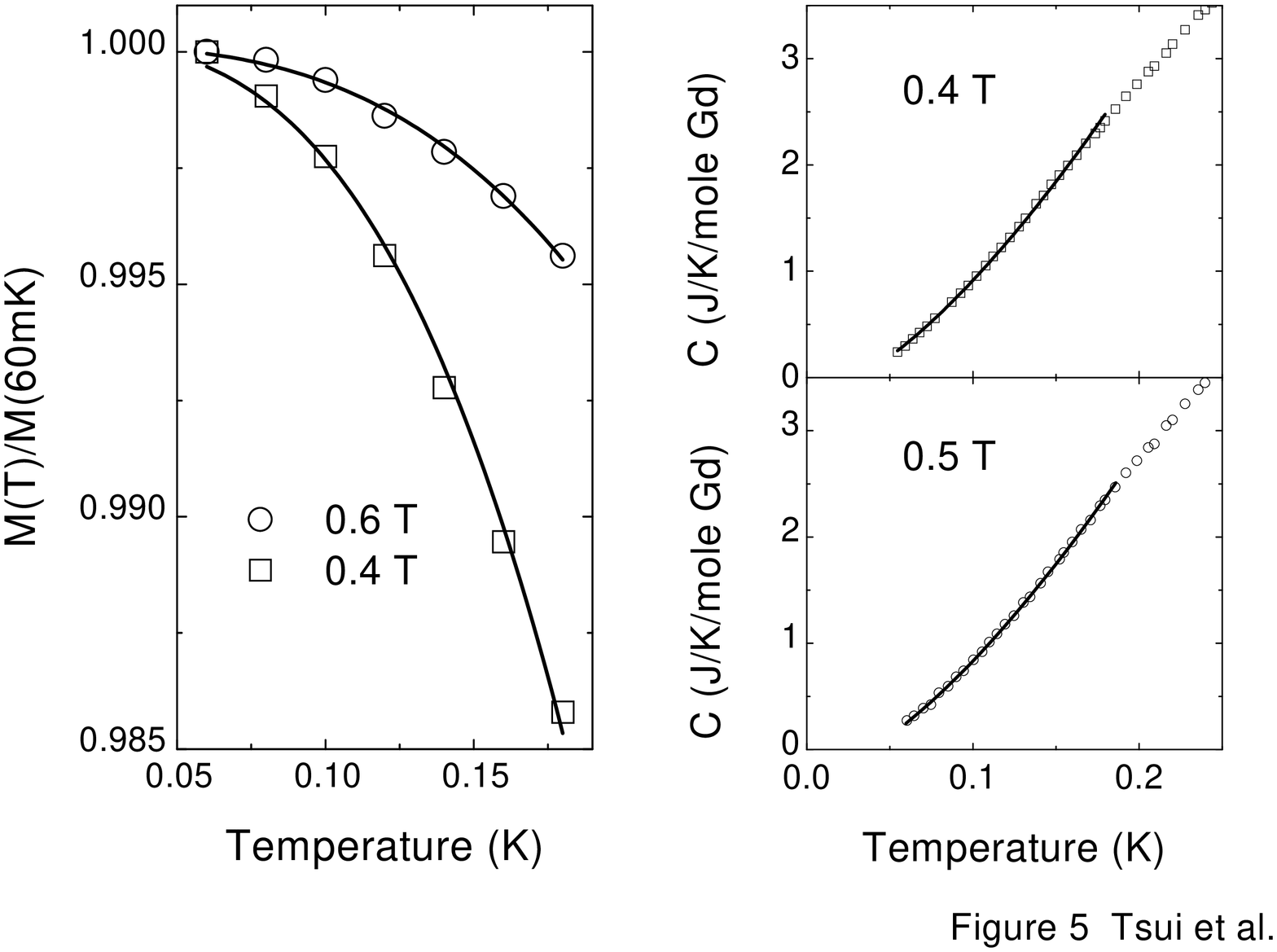}
\vspace{1in}
\caption{Left: Low-temperature $M$ versus temperature at 0.4 and 0.6 T. The
solid
lines are fits to the data as described in the text. Right: Low-temperature $C_p$
versus
temperature at 0.4 and 0.5 T respectively. The solid lines are fits to
the data as described in the text.}
\label{fig:CFit}
\end{figure}

To further investigate the nature of the low temperature excitations in the spin liquid 
phase, we closely examined our $C(T)$ data for T$ < T_{cross}$ within the spin liquid 
regime. Since $C(H)$ shows ferromagnetic-like behavior (i.e. $\partial C(H)/\partial H < 0$), we fit the low temperature $C(T)$ data to a spin-wave-like model with the dispersion relation, $\hbar\omega = \Delta + Dk^2$, where $D$ is the spin stiffness constant and $\Delta$ is an energy gap.  Note that while this dispersion relation has the same form as that for ferromagnetic spin waves, in fact we are only assuming that the excitations are non-interacting bosons with a gapped quadratic dispersion relation. In this model, we calculate the specific heat as

\begin{equation} \label{eqn:CX0}
C_{\mathrm{model}}(T) = 
       \frac{k_{\mathrm{B}}^{5/2}T^{3/2}}{4\pi^2D^{3/2}}
       \int_{x_0}^{\infty} \frac{x^2\sqrt{x-x_0}\exp(x)}{(\exp(x)
-1)^2}dx~,
\end{equation}
where $x_0 = \Delta/k_{\mathrm{B}}T$.  As shown in figure \ref{fig:CFit}, equation \ref{eqn:CX0} fits the measured $C(T)$ data well \cite{caveat}, and we obtain $D \sim$ 2 x $10^{-43} Jm^2$ which is almost independent of 
field and $\Delta$ which is linear in field as shown in the inset to figure \ref{fig:CT}. Moreover, since

\begin{equation} \label{eqn:SandM}
  \left(\frac{\partial S}{\partial H}\right)_T = 
  \left(\frac{\partial M}{\partial T}\right)_H~,
\end{equation}
we can get the temperature dependence of $M$ by integrating the expression for the specific heat $C$,

\begin{equation} \label{eqn:MX0}
M(T) \sim \int_0^T dT^{\prime}
\int_0^{T^\prime} \frac{C(T^{\prime\prime})}{T^{\prime\prime}}dT^
{\prime\prime}~.
\end{equation}
We perform this integral numerically, substituting equation \ref{eqn:CX0}
in equation \ref{eqn:MX0}, and we find that the numerical form for the magnetization can be described by

\begin{equation}\label{eqn:MT}
M(T) \sim M_0(1 - AT^{3}~),
\end{equation}
in the temperature range where we fit $C(T)$ ($60 \lesssim T \lesssim 200$ mK). As shown in figure \ref{fig:CFit}, equation \ref{eqn:MT} does fit the measured $M(T)$ very well at low 
temperatures, lending further credence to the model.

Since the model apparently describes the measured values of $C(T)$ and $M(T)$ quite well, we can examine the obtained fit parameters to obtain some understanding of the nature of the excitations.  The fitted values of the spin stiffness parameter, $D$, vary by less than 10\% in our range of fields.  If we take the usual assumption for a ferromagnet that the ferromagnetic exchange energy $J = D/2Sa^2$ where S = 7/2 for the Gd ions and $a$ = 3 \AA ~is the nearest neighbor Gd separation, we obtain J $\sim 0.02 K$ which 
corresponds to the expected weak ferromagnetic correlation in the canted spin liquid. 
 
	One of the most striking features of the results is the presence of a gap in the spin excitations, without which the data could not be fit with an integer power law dispersion relation. Corroborating evidence for the existence of such a gap can be found in the earlier a.c. susceptibility data which showed that $\chi_{ac}$ decreases with decreasing temperature in the spin liquid phase \cite{Schi94,Schi00}. Further physical justification for the existence of the gap can be seen in the inset to figure \ref{fig:CT}, which shows that that the fitting parameter $\Delta$(H) is linear in 
field as would be expected for an energy gap.  We find that $\Delta$(H) can be fit well by $\Delta$(H) = $\Delta_0 - \mu_{exc}$H, and we associate $\mu_{exc}$ = -0.16 $\mu_B$ with a negative effective moment of the spin excitations which is much smaller than the moment of the S = 7/2 Gd spins.  This finding is consistent with the fundamental magnetic unit of the spin liquid 
being a low-moment correlated cluster of spins in the spin liquid as has been suggested theoretically \cite{clusters,Spinliqtheor}.  The extrapolated zero-field gap, $\Delta_0$/$k_B \sim$ 0.03K, could be attributable an intrinsic gap in the spin liquid or the anisotropy associated with the relatively large dipole interactions between the Gd spins.  Our value of $\Delta_0$/$ k_B \sim  \Theta_{\mathrm{cw}}/70$ is somewhat smaller than that predicted theoretically for the intrinsic gap in a kagom\'{e} system \cite{Spinliqtheor}, but it is of the same order.  If we attribute $\Delta_0$ instead to a finite-size effect, we can estimate the correlation length of the excitations,
$L_0 \sim \sqrt{4\pi^2D/\Delta_0} \sim 40$ \AA~.  This length scale is comparable to the $\sim$ 100 \AA~ correlation length observed in recent zero-field neutron scattering studies \cite{Petr97}.  Interestingly, despite the first order transition between the spin liquid and the long-range-ordered antiferromagnetic phase at higher fields, the same functional form can fit C(T) in the long-range-ordered phase at low temperatures with a somewhat larger value of $D$, and $\Delta_0 \sim  0$.  This suggests both that $\Delta_0$ originates from a finite-size effect in the spin liquid phase and that the short range antiferromagnetic order in the spin liquid phase is of the same sort as that in the long-range-ordered phase.

In summary, we have investigated the low temperature excitations of a 
three dimensional spin liquid through its thermodynamic properties. We find that there is a 
crossover temperature below which the thermal fluctuations change character, which is apparently a generic feature of geometrically frustrated 
spin liquid states.  We also find that we can describe the magnetization and 
heat capacity data by a simple model in which the excitations have a gapped quadratic dispersion relation analogous to ferromagnetic spin waves.  These data provide the first thermodynamic evidence for the nature of the excitation spectrum in a three dimensional spin liquid phase \cite{Gaulin}, and they indicate the need for new single-crystal inelastic neutron scattering investigations of these systems.

The authors acknowledge the financial support of the NSF grant
DMR97-01548, ARO grant DAAG55-98-1-0032, and the Alfred P. Sloan Foundation.
We are grateful for the helpful discussion with A. P. Ramirez and D. A.
Huse.  We thank V. Fratello and A. J. Valentino at Bell Laboratories 
for sample preparation, and N. Kalechofsky (now at Oxford Instruments) for
assistance in the early stages of these experiments.

\newpage
\vspace{1in}



\begin{references}

\bibitem{1dliq}
See, for example G. Y. Xu et al., Science {\bf 289}, 419 (2000) and references therein.

\bibitem{cupliq}
See, for example W. J. Zheng, Phys. Rev. Lett. {\bf 83}, 3534 (1999) and references therein.

 \bibitem{Rami94}
A.P. Ramirez, Annu. Rev. Mater. Sci. {\bf 24}, 453 (1994);
P. Schiffer and A.P. Ramirez,
Comments on Condensed Matter Physics {\bf 18}, 21 (1996);
M.J. Harris and M.P. Zinkin,
Mod. Phys. Lett. {\bf B 10}, 417 (1996).


\bibitem{clusters}E. F. Shender et al., Phys. Rev. Lett.{\bf 70}, 3812 (1993); P. Schiffer and I. Daruka, Phys. Rev. B {\bf 56}, 13712 (1997); R. Moessner and A.J. Berlinsky, Phys. Rev. Lett. {\bf 83}, 3293 (1999), P. Mendels {\it et al}., Phys. Rev. Lett. {\bf 85}, 3496 (2000).

\bibitem{Spinliqtheor}See, for example,
R. Moessner and J.T. Chalker, Phys. Rev. Lett. {\bf 80}, 2929 (1998);
B. Canals and C. Lacroix, Phys. Rev. Lett. {\bf 80}, 2933 (1998) and Phys. Rev. B {\bf 61}, 1149 (2000);P. Sindzingre {\it et al}., Phys. Rev. Lett. {\bf 84}, 2953 (2000) and references therein.

\bibitem{Rami90}
A.P. Ramirez {\it et al.}, Phys. Rev. Lett. {\bf 64}, 2070 (1990);
Phys. Rev. {\bf B 45}, 2505 (1992);
Y.J. Uemura {\it et al.}, Phys. Rev. Lett. {\bf 73}, 3306 (1994);
C. Broholm {\it et al.}, Phys. Rev. Lett. {\bf 65}, 3173 (1990).

\bibitem{Schi94}
P. Schiffer {\it et al.}, Phys. Rev. Lett. {\bf 73}, 2500 (1994) and Phys. Rev. Lett. {\bf 74}, 2379 (1995).

\bibitem{harris}
M. J. Harris {\it et al.}, Phys. Rev. Lett. {\bf 73}, 189 (1994), Phys. Rev. B {\bf 52}, R707 (1995).

\bibitem{Petr97}
O.A. Petrenko {\it et al.},
Physica B {\bf 241-243}, 727 (1997);
Phys. Rev. Lett. {\bf 80}, 4570 (1998);  Physica B {\bf 266}, 41 (1999).

\bibitem{Ball96}
R. Ballou, E. Leli\`{e}vre-Berna, and B. F\aa k,
Phys. Rev. Lett. {\bf 76}, 2125 (1996).

\bibitem{Tsui99}
Y.K. Tsui {\it et al.}, J. Appl. Phys. {\bf 85}, 4512 (1999);
Phys. Rev. Lett. {\bf 82}, 3532 (1999);
Physica B {\bf 280}, 296 (2000); Y.K. Tsui, PhD. thesis, University
of Notre Dame (unpublished).

\bibitem{Gard99}
J.S. Gardner {\it et al.}, Phys. Rev. Lett. {\bf 82}, 1012 (1999)

\bibitem{RaHeWi00} A.P. Ramirez, B. Hessen, and M. Winklemann,
Phys. Rev. Lett. {\bf 84}, 2957 (2000).

\bibitem{Duns00} S. R. Dunsinger et al.,
Phys. Rev. Lett. {\bf 85}, 3504 (2000).

\bibitem{shiga} 
M Shiga and H Nakamura J. Phys. Soc. Jap. {\bf 69}, 147 (2000) and references therein.

\bibitem{KiWo79}
W.I. Kinney and W.P. Wolf,
J. Appl. Phys. {\bf 50}, 2115 (1979);
W.I. Kinney. Ph.D. thesis, Yale University, 1979 (unpublished).
\bibitem{Wolf62} 
W.P. Wolf {\it et al.}, J. Phys. Soc. Jpn. B1 
{\bf 17}, 443 (1962);
D.G. Onn, H. Meyer, and J.P. Remeika,
Phys. Rev. {\bf 156}, 663 (1967);
R.A. Fisher {\it et al.},
J. Chem. Phys. {\bf 59}, 4652 (1973).


\bibitem{Hov80} S. Hov, H. Bratsberg, and A.T. Skjeltorp,
J. Magn. Magn. Mater. {\bf 15-18}, 455 (1980); S. Hov, Ph.D. thesis,
University of Oslo, 1979 (unpublished);A.P. Ramirez and R.N. Kleiman,
J. Appl. Phys. {\bf 69}, 5252 (1991).

\bibitem{Saka94}
Toshiro Sakakibara {\it et al}., Jpn. J. Appl. Phys. {\bf 33}, 5067
(1994).


\bibitem{Mydosh93} J.A. Mydosh {\it Spin glasses: an experimental
introduction},
(London, Taylor \& Francis, 1993).

\bibitem{energyscale} Because of the lower energy scale of the interactions, 
the field dependence is, in fact, comparatively much weaker in GGG than in SrCr$_{\mathrm{9p}}$Ga$_{\mathrm{12-9p}}$O$_{19}$.
 
\bibitem{caveat}
Our model only fits the data for fields $\gtrsim$ 0.3 T in the spin liquid regime.  This is probably associated with the smaller magnetization at lower fields which would not support a propagating spin wave.

\bibitem{Schi00}
P. Schiffer, unpublished.

\bibitem{Gaulin} B. D. Gaulin, private communication.

\end{references}
\end{document}